\documentclass[aps,prd,twocolumn,groupedaddress,nofootinbib]{revtex4-1}
\usepackage{graphicx}
\usepackage{amsmath}
\usepackage{color}
\usepackage{ulem}
\usepackage{hyperref}

\hypersetup{
	colorlinks=true,       
	linkcolor=blue,          
	citecolor=blue,        
	urlcolor=blue           
}

\begin{document}
	
	
\title{Constraints on Neutron-Mirror-Neutron Oscillation from Neutron Star Cooling}
	
	\author{ Itzhak Goldman$^{a,b}$}
	
	\author{ Rabindra N. Mohapatra$^c$}
	
	\author{ Shmuel Nussinov$^b$}
	
	\author{ Yongchao Zhang$^d$}
	\affiliation{$^a$ Afeka College, Israel}
	\affiliation{$^b$ Tel Aviv University, Israel}
	\affiliation{$^c$ Maryland Center for Fundamental Physics and Department of Physics, University of Maryland, College Park, Maryland 20742, USA}
	\affiliation{$^d$ School of Physics, Southeast University, Nanjing 211189, China}
	
\begin{abstract}
		

  We address a method of limiting neutron-mirror neutron mixing ($\epsilon_{nn'}$) by analyzing its effect on neutron star (NS) heating. This method employs observational bounds on the surface temperature of NSs to constrain $\epsilon_{nn'}$. It has been suggested that the bound obtained this way is so stringent that it would exclude any discovery of $n-n'$ oscillation in the currently planned terrestrial experiments at various laboratories. This conclusion motivated us to critically analyze this suggestion in more detail. In this note, we point out a very interesting new effect present in nearly exact mirror models, which can significantly affect this bound. The new element is that in nearly exact mirror models there is the mirror analog of $\beta$ decay, i.e. $n' \to p' + e' + \bar\nu'_e$, which creates a cloud of mirror particles $n'$, $p'$, $e'$, $D'$ and He$'$ inside the NS. The resulting $e'$  can ``rob'' the energy generated by the $n \to n'$ transition from the NS, via $e-e'$ scattering enabled by the presence of a (minute) millicharge in mirror particles. Such a tiny millicharge on mirror particles is highly likely in these models. This results in energy being emitted as unobserved mirror photons via fast mirror bremsstrahlung, whose effect is to relax the stringent bounds on $\epsilon_{nn'}$.
		
	\end{abstract}
	
	\date{\today}
	
	\maketitle
	
	\section{Introduction}
	
	
	Neutron stars  (NSs)  and their violent supernova birth events have played a special role in constraining physics beyond the standard model (BSM)~\cite{Raffelt:1996wa}. This is especially true for BSM scenarios where new light particles such as axions, majorons and light sterile neutrinos coupling to standard model (SM) particles are present. Most constraints arise from energy loss arguments in the supernova explosion. Another type of signatures arises in a class of BSM scenarios called mirror models, where the heating of a NS becomes the source of constraints.  We focus here on this latter class of models.
	
The mirror models contain an identical duplicate of the particles and forces of the SM, coexisting with it in the same universe, but invisible to us since all its forces except gravity are different from ours. These models were proposed by Lee and Yang in their parity violation paper in order to maintain parity as a good symmetry of nature despite the maximal parity violation in observed weak interactions~\cite{Lee:1956qn,Kobzarev:1966qya,Blinnikov:1983gh,Foot:1991bp, Hodges:1993yb, Foot:1995pa, Berezhiani:1995yi, Berezhiani:2000gw, Ignatiev:2003js,Berezhiani:2003wj, Ciarcelluti:2004ik, Ciarcelluti:2004ip, Sandin:2008db, Ciarcelluti:2010zz, Foot:2014mia}. Under parity the particles and forces of the SM transform to those of the mirror SM, and the cross-interactions between the particles in two sectors are dictated by a priori unknown couplings, e.g. the kinetic mixing of photon with the mirror photon~\cite{Berezhiani:2003xm, Berezhiani:2008zza}. They have been extensively studied in the past three decades, in connection with dark matter (DM) possibly residing in the mirror sector as well as sterile neutrinos being the mirror partners of the familiar ones. In a special subclass of these models, where the mirror symmetry is almost exact, there arises the possibility of a highly degenerate neutron-mirror neutron system, raising the possibility of neutrons oscillating to mirror neutrons ($n-n'$ oscillation)~\cite{Berezhiani:2005hv}.

In general, the Hamiltonian for the $n-n'$ system can be written in the form of
\begin{eqnarray}
\label{eqn:Hamiltonian}
{\cal H} = \begin{pmatrix}
m_n + \Delta E & \epsilon_{nn'} \\ \epsilon_{nn'} & m_{n'}
\end{pmatrix} \,,
\end{eqnarray}
with $m_n$ and $m_{n'}$, respectively, the masses of $n$ and $n'$, $\epsilon_{nn'}$ the parameter enabling $n-n'$ mixing.  There are generally two (small) parameters that are important for the $n-n'$ transition, i.e. the off-diagonal mass parameter $\epsilon_{nn'}$ and the mass splitting $\delta_{nn'} \equiv |m_{n'} - m_n|$. In addition, there is another parameter $\Delta E$, denoting the environmental effects contributing to the mass splitting in the matrix (\ref{eqn:Hamiltonian}), which can arise for instance due to mirror gas density or mirror magnetic fields. This parameter is essentially an unknown parameter and introduces an inherent uncertainty into the conclusion about $\epsilon_{nn'}$ derived from an experiment. Thus from any laboratory search, one can get a broad range of limits on $\epsilon_{nn'}$, depending on assumptions about $\Delta E$. This is the case for exact mirror models.

On the other hand, if we take $\delta_{nn'}\neq 0$, a further parameter is introduced. This could happen in asymmetric mirror models. If $n-n'$ transition is searched in a non-zero magnetic field $B$, it could in principle compensate the effect of $\delta_{nn'}$ by the equivalent energy $\Delta E = \mu_n B$ induced by magnetic field, with $\mu_n$ being the neutron magnetic moment~\cite{Berezhiani:2009ldq}. It is even possible that magnetic field of a proper strength can resonantly enhance the $n-n'$ transition. But, again due to our lack of knowledge, we will never know for sure about the mirror magnetic field. Nevertheless, it is important to carry out the searches for $n-n'$ oscillation in the hope that some such possibility can occur and we will discover the $n-n'$ oscillation. A discovery of $n-n'$ oscillation will, however, provide us another input to determine $\epsilon_{nn'}$ in a certain range. 

Thus to summarize this part, if the $n-n'$ transition exists with a time scale $\tau_{nn'} \equiv 1/ \epsilon_{nn'}$ at the order of ten seconds, it can be detected in a rather simple disappearance $n \to n'$ and/or regeneration $n \to n' \to n$ type experiment with
an intense beam of cold neutrons, provided the mirror magnetic field is not too large~\cite{Pokotilovski:2006gq, Berezhiani:2017azg, Kamyshkov:2021kzi}.
The current existing laboratory limits on $n-n'$ mixing are from Refs.~\cite{Ban:2007tp, Serebrov:2007gw, Altarev:2009tg, Serebrov:2008her, Berezhiani:2012rq, Berezhiani:2017jkn, nEDM:2020ekj, Mohanmurthy:2022dbt, Ayres:2021zbh, Broussard:2021eyr}. The experimental results are conventionally presented in terms of the time scale $\tau_{nn'}$. In particular, the strongest limit among them is from Ref.~\cite{Serebrov:2007gw}, which excludes the transition time scale $\tau_{nn'} < 414$ sec, corresponding to the limit of $\epsilon_{nn'} < 1.6 \times 10^{-18}$ eV assuming exact mirror models so that $\delta_{nn'}=0$ and the effect of the mirror magnetic field is somehow compensated or suppressed. In other words, such a limit applies only to the small splitting $\delta_{nn'} < 10^{-14}$ eV and $\Delta E=0$  and becomes invalid for a larger $\Delta E$. The combined results of other experiments searching for $n-n'$ transitions at different magnetic fields still allow $\epsilon_{nn'}$ as small as $3\times 10^{-17}$ eV in the range of mass splitting up to $\delta_{nn'}\simeq 10^{-12}$ eV. For larger splitting, the experimental limits sharply drop (see Fig.7 in Ref.~\cite{Berezhiani:2017jkn}): for say $\delta_{nn'} > 10^{-13}$ eV, they allow $\epsilon_{nn'} > 10^{-15}$ eV~\cite{Berezhiani:2005hv}.
Some of these experiments have observed anomalies up to $5\sigma$,  which could imply $n-n'$ oscillation with $\epsilon_{nn'}= (1\div 2) \times 10^{-17}$ eV~\cite{Ban:2007tp, Serebrov:2008her, Berezhiani:2012rq, Berezhiani:2017jkn}. Therefore, experiment was designed to test these anomalies, which can exclude the value of $\epsilon_{nn'}$ down to $5 \times 10^{-18}$ eV for the range of $\delta_{nn'}$ up to few times $10^{-11}$ eV~\cite{Ayres:2021zbh}. The sensitivities can be further improved at the European Spallation Source~\cite{Addazi:2020nlz} and other experiments~\cite{Broussard:2021eyr}.

Another result in neutron physics that motivates the $n-n'$ search is the neutron lifetime anomaly. The lifetime of neutron can be determined by observing the disappearance of ultra-cold neutrons in magnetic traps~\cite{Mampe:1993an, Serebrov:2004zf, Serebrov:2007ve, Pichlmaier:2010zz, Steyerl:2012zz, Arzumanov:2015tea, Serebrov:2017bzo, Ezhov:2014tna, Pattie:2017vsj}, or from the appearance of decay products (protons)  in a beam experiment~\cite{Wietfeldt:2011suo, Serebrov:2011re, Greene:2016}. However, there has been tension between the measured neutron lifetime in these two methods, at the level of $4\sigma$~\cite{Byrne:1996zz, Nico:2004ie, Yue:2013qrc}. Such a discrepancy can be explained by the $n-n'$ oscillation, in the parameter range of $\delta_{nn'} > 3 \times 10^{-7}$ eV and $\epsilon_{nn'} / \delta_{nn'} > 10^{-3}$ or so~\cite{Berezhiani:2018eds, Berezhiani:2018udo}. Some regions of this parameter space have been excluded by the limit from Ref.~\cite{Broussard:2021eyr}. In fact, the $n-n'$ mass splitting could be as large as $\delta_{nn'} \simeq 1$ MeV, in which case the neutron lifetime anomaly can be tackled via $n \to n'$ decay as in Ref.~\cite{Berezhiani:2005hv}.
	
The $n-n'$ oscillation with small oscillation times, in any case smaller than the neutron decay time of roughly 880 sec, can have interesting astrophysical implications~\cite{Mohapatra:2005ng, Berezhiani:2006je, Berezhiani:2011da}.
	Since  NSs are extremely rich in neutrons, they  are  perfect laboratories for testing the implications of $n-n'$ oscillations.
	{The transition of an ordinary neutron $n$ to a mirror neutron $n'$ is followed by a drop of the latter towards the NS center  (under gravity). The hole left will then be filled by another neutron $n$, and in the process the NS will lose part of the mass and  energy will be liberated~\cite{Berezhiani:2020zck, Mannarelli:2018}}. If the process is fast enough, it would lead eventually to a fully mixed star~\cite{Ciarcelluti:2010ji}. The mass loss  of a NS, whatever the reason for it is, can manifest in many ways: for instance it can lead to changes in the orbital period of a binary pulsar, regardless of the mechanism by which mass loss occurs~\cite{Goldman:2009th}.  Using observational constraints on the rate of the binary periods for several binary pulsars, upper bounds can be set on the corresponding $n-n'$ mixing parameter $\epsilon_{n n'}$. The limits on the period change in several pulsars led to an upper limit of $10^{-13}$ eV  on $\epsilon_{nn'}$,  which  is valid for mass splitting $\delta_{nn'}$ up to few MeV~\cite{Goldman:2019dbq}.\footnote{If $n'$ is lighter than $n$, the $n - n'$ mass splitting can be at most roughly few MeV from nuclear stability~\cite{Berezhiani:2005hv}. In less dense regions of NSs, the matter induced splitting $\delta_{nn'}$ can be less than 15 MeV, depending on equation of state.} Furthermore, the NS arguments are independent of $\delta_{nn'}$,  unless the mass splitting is very large and also the magnitude of the mirror magnetic field, which makes the NS limits universal and interesting.
	
More recently, it was noted that there can be another drastic effect of $n\to n'$ transition in a single pulsar as well~\cite{Berezhiani:2020zck, McKeen:2021jbh}. It was well known that when a neutron converts to a mirror neutron, the hole left  by the disappearing neutron is quickly filled by a neutron near its Fermi level. In this process a considerable amount of energy can be released, which will affect the luminosity of the NS. Taking the coldest NS,  PSR J2144$-$3933~\cite{Guillot:2019ugf}, it was  found that the $n - n'$ mixing parameter $\epsilon_{n n'} \lesssim 10^{-17}$ eV~\cite{Berezhiani:2020zck, McKeen:2021jbh}. This bound  is  valid for $n-n'$ mass difference $\delta_{nn'}$ up to 10 MeV~\cite{McKeen:2021jbh}.
However, the relation between the $n-n'$ transformation and $\epsilon_{nn'}$ is highly non-trivial, depending on the nuclear model, the NS mass etc. In Ref.~\cite{Berezhiani:2020zck}, the limit on $\epsilon_{nn'}$ is estimated from general arguments (see more concise estimates in Ref.~\cite{Berezhiani:2021src}), while in Ref.~\cite{McKeen:2021jbh} it was numerically calculated for one particular equation of state, with a considerably large uncertainties in the final result. Therefore, the limit of $\epsilon_{nn'} \lesssim 10^{-17}$ eV can only be considered as an order of magnitude estimate, within a factor of few.
This bound is particularly important, since currently planned terrestrial experiments are sensitive to $\epsilon_{nn'}$ at the level of $10^{-17} $ eV~\cite{Ban:2007tp,Serebrov:2007gw, Altarev:2009tg, Serebrov:2008her, Berezhiani:2012rq, Berezhiani:2017jkn, nEDM:2020ekj, Mohanmurthy:2022dbt, Ayres:2021zbh, Broussard:2021eyr}. However, it has been pointed out in our previous paper that there is a loophole in this argument~\cite{Goldman:2022rth}, and the present paper is an elaboration of this result.



	
	It is well known that if the energy emitted from astrophysical objects is even partially in the form of  electromagnetic energy, then the expected signal and ensuing bounds would be dramatically enhanced. The works of Refs.~\cite{Dar:1987nq,Kazanas:2014mca,Hook:2018iia} are a few examples among many related to supernovae and NSs, and the NS conversion into mixed star is no exception.
	Ref.~\cite{McKeen:2021jbh} used the fact that the heat generated in old, cold pulsars is emitted directly via photons~\cite{Yakovlev:2004iq, Potekhin:2020ttj}. This leads to  a strong upper bound on  the rate $\Gamma_{n\to n'}$ of neutron to mirror neutron conversion, which translates to an upper bound on the mixing parameter  $\epsilon_{nn'}\leq 10^{-17}$  eV.
	
	In this paper we critically analyze this  by following the evolution of the $n'$ generated in $n-n'$ transition a bit longer. We use the fact that in almost exact mirror models, the  mirror neutrons generated inside the NS $\beta$ decay producing mirror electrons $e'$, mirror protons $p'$ and mirror neutrinos $\bar{\nu}'_e$. These mirror charged particles then provide a competing cooling channel via the emission of mirror photons, and reduce the photonic signal claimed in Ref.~\cite{McKeen:2021jbh}, considerably relaxing   the upper bounds on $\epsilon_{nn'}$. While the strong gravity in the NS causes the mirror particles to sink towards the center of the star, we find that for a relatively wide acceptable range of interactions between the ordinary and mirror sectors, mediated by the millicharge of mirror particles, the  nucleons and electrons of the visible sector in this core region of the NS can transfer their energy to the mirror particles. The latter then emit this energy via mirror photons, which do not interact with the ordinary nucleons and electrons and can freely escape.
	
	The plan of this paper is as follows: in Section~\ref{sec:conversion} we briefly recap the thermal state of the NS in the absence of any mirror particles, and describe the first steps in the slow $n-n'$ conversion. The following sections  contain  the  new result of the present work. We first show in Section~\ref{sec:decay} that in the case of $m_n\simeq m_{n'}$ the $ \beta$ decay $n'\to p' + e' +\bar{\nu}'_e$ occurs in the NS. This is followed by the fast electromagnetic capture process  $n'+p'\to D' + \gamma'$ and the formation of other heavier nuclei $X'$. Next in Section~\ref{sec:profile} we address the hydrostatics of the mirror $e'-X'$ fluid, which is dominated by the stellar gravity operating on the $X'$s and the Fermi pressure of the light $e'$s.  We then  describe in Section~\ref{sec:energy} the sort of ``mini white dwarf'' formed  inside the NS, and indicate how  the small {milli-charge $\epsilon$} leads to dramatic cooling effects of the star for non-negligible non-gravitational couplings between the two sectors.  Following that, in Section~\ref{sec:comments} we briefly comment on the possible impact of ultra cold neutron stars (UCNSs) on a variety of new physical effects. In Section~\ref{sec:accretion} we note that searching for such UCNS, as providing signatures for new BSM physics, is not hampered by accretion of interstellar medium (ISM) gas.  We also comment  on the apparent uniqueness of the coldest pulsar  PSR J2144$-$3933, which is used to derive the strongest upper bounds on $\epsilon_{nn'}$. We summarize in Section~\ref{sec:summary}. The details of the cross section for $e - e'$ scattering are given in Appendix~\ref{sec:xs}.
	
	
	\section{The Neutron star PSR 2144-3933 and $n-n'$ conversion rate }
	\label{sec:conversion}
	
	We now review  some of the basic information about the pulsars given in Refs.~\cite{Yakovlev:2004iq, Potekhin:2020ttj}, which give  a listing of 55 different pulsars of various ages, magnetic fields, surface temperatures and luminosities. Initially, shortly after their birth, the NSs are relatively hot and they cool down via volume emission of neutrino pairs. Eventually its internal temperature  $T_{\rm int}$ drops, and the neutrino luminosity which scales as $T_{\rm int}^8$ becomes negligible. The goal of Refs.~\cite{Yakovlev:2004iq, Potekhin:2020ttj} was to understand the heat conduction and emission mechanism of such electromagnetically emitting pulsars.
	At the time of observation, the star may be still cooling off or, if some other sources of energy
	exist, it may have settled into a thermal steady state, with the thermal energy emitted
	as electromagnetic radiation often as a black body radiation. Let us apply this scenario to the pulsar PSR J2144$-$3933.
	
	In a steady state, the black body luminosity (the power radiated from the star surface) is given by the Stefan-Boltzmann formula:
	\begin{eqnarray}
		\label{eqn:luminosity}
		\frac{{\rm d}W}{{\rm d}t} = {\cal L}_{\rm NS} = 4\pi \sigma_{\rm SB} R^2 T_s^4 \,,
	\end{eqnarray}
	where $\sigma_{\rm SB}$ is the Stefan-Boltzmann constant, $R$ is the radius of the NS, and its external surface temperature $T_s$ is maintained by the constant internal energy source. If we have observational limits on the luminosity, this implies upper bounds on the rate of internal heat production.
	
	Finding stars with black body radiation would of course not prove the scenario of $n-n'$ transition and resulting mass reduction of NS, as it could reflect, for example, the ongoing cooling of the initial hot star, axion conversion in the magnetic fields, accretion of DM, accretion of ISM gas, or any other activity. However, finding sufficiently cold  pulsars, for which strong bounds on their electromagnetic emissions are available, will provide independent upper bounds on every potential heating mechanism. The colder the pulsar considered, the lower its luminosity is and the stronger is the resulting bound.
	
	In discussing constraints on the internal heat in a NS, it is important to note the feature
	that there is a $\sim$100 meter thick nuclear ``thermal blanket'' just under the surface~\cite{Beznogov:2021ijc}. It causes the internal temperature, which is almost uniform over the NS, to drop dramatically by a factor of  $\sim$100  as we move out from the inside across the blanket towards the surface. The estimated upper bound on surface temperature
	$T_s\sim 42000$ K of the coldest pulsar PSR J2144$-$3933 \cite{Guillot:2019ugf} would  then corresponds to $ T_{\rm int}  \sim 4.2 \times10^{6}$ K $\simeq 0.35$ keV. This internal temperature  would play an important role in obtaining upper bounds on any heat generating mechanism.
	
	If the $n \to n'$ processes were the only source of heat supply, then in a steady state the overall $n-n'$ transition rate would be given by
	\begin{eqnarray}
		\frac{{\rm d}{\cal N}_{n'}}{{\rm d}t} = \frac{{\cal L}_{\rm NS}}{\Delta E}  \,,
	\end{eqnarray}
	where $\Delta E$ is the energy initially gained by ordinary nucleons in each $n\to n'$ transition. In degenerate neutron matter, the Fermi momentum is about $(n_N/0.15 \, {\rm fm}^{-3})^{1/3} \times (300\, {\rm MeV})$, with $n_N$ the nucleon number density ($N =p,\; n$ is nucleon in the star). It is expected that only a small fraction of neutrons close to the Fermi surface can take part in the $n \to n'$ transition, and the mirror neutron $n'$ takes away a sizable fraction of the total momentum of nucleons in the process $n N \to n' N$. To be concrete, we take explicitly $\Delta E = 30$ MeV for the calculations below. However, one should note that, this is only an order of magnitude estimate, and the factors of few can make a difference (see e.g. Ref.~\cite{Berezhiani:2020zck}).

For PSR J2144$-$3933, taking  $R=12$ km, the luminosity inferred from Eq~(\ref{eqn:luminosity})
	\begin{eqnarray}
		\label{eqn:LNS}
		{\cal L}_{\rm NS} \simeq 3\times10^{27} \ {\rm erg/sec}
	\end{eqnarray}
	then translates into the rate of generating new mirror neutrons:
	\begin{eqnarray}
		\label{eqn:dNpdt}
		\frac{d {\cal N}_{n'}}{dt} \sim 0.5\times 10^{32} \left( \frac{T_s}{42000\ {\rm K}} \right)^4 \; {\rm sec}^{-1}
	\end{eqnarray}
	Some pulsars in the sample of Refs.~\cite{Yakovlev:2004iq, Potekhin:2020ttj} have temperatures up to 100 times higher yielding ${\rm d}{\cal N}_{n'}/{\rm d}{t} \sim 10^{40}$ sec$^{-1}$, and were also used to bound high $\epsilon_{nn'}$ values.  Returning to our oldest  PSR J2144$-$3933, we find from Eq.~(\ref{eqn:dNpdt}) that during its long lifetime of 330 million years  about
	\begin{eqnarray}
		{\cal N}_{n'} \sim 10^{48}
	\end{eqnarray}
	neutrons would convert into mirror neutrons. This comprises a tiny ${\cal N}_{n'}/{\cal N}_{n} \sim 10^{-9}$  fraction of the total nucleon number in the star, with no change of the gravity fields and of the local density profile of the ordinary NS.
	
	Backtracking a bit, we recall the basic energetics of the $n\to n'$ transitions inside the NS. This underlies their robustness and that of the resulting upper bounds on $ \epsilon_{nn'}$  of the $n-n'$ mass difference.  Unlike the much earlier suggested $n-\bar{n}$ oscillations~\cite{Kuzmin:1970nx, Glashow:1979nm, Mohapatra:1980qe}, the $n-n'$ transitions cannot happen in nuclei, as such transitions will lower their energy by the $n'$ binding of $\sim 8$ MeV. In the NS, however, nuclear binding effects are secondary to gravity. The newly born $n'$ escapes from its original location carrying along its initial $\sim 30$ MeV Fermi energy, yet the strong gravity prevents it from escaping the NS.
	
	Neighboring neutrons rush into the ``hole'' formed, and the work done in the process is
	\begin{eqnarray}
		{\rm d}W = p(r) {\rm d} V \,,
	\end{eqnarray}
	where $p(r)$ is the local pressure, and $ {\rm d}V \sim$ fm$^3$ is the volume of the hole. The work indicated in the equation above is also $\sim 30$ MeV on average and becomes ``heat'', namely kinetic energy of these nucleons. These nucleons collide with neighboring neutrons with density $ n_{N} \sim 10^{39}$ cm$^{-3}$, and very quickly settle into the spatially and temporally fixed internal
	temperature $T_{\rm int}$ ($ \sim  0.35 $ keV).
	
	Finally we note that only the $kT/E_{F}$ fraction of nucleons and electrons in the high energy tail of the   degenerate Fermi-Dirac energy distribution are not Pauli blocked and can be excited (or de-excited) to higher (or lower) empty energy states, reducing by $kT/{E_F}  $ the specific heat and the heat content $Q^*$ of the NS.  It is then given by
	\begin{eqnarray}
		\label{eqn:Qstar}
		Q^*=  \frac{{\cal N}_n (k T)^2}{E_F}  \,.
	\end{eqnarray}
	Upon using $kT \sim 0.35$ keV appropriate to  PSR J2144$-$3933, and $E_F = 30$ MeV (so that  $kT/E_F\sim 10^{-5}$), and the total number of nucleons in the star ${ \cal N}_{n}  \sim 2\times 10^{57}$, we find
	\begin{eqnarray}
		Q^* \sim   10^{52} \ {\rm keV} \,.
	\end{eqnarray}
	The fact that only the $kT/E_F$ fraction of these end point ``active'' electrons will partake in electron scattering or any other dynamic processes, will play an important role in the following  calculations.
	
	
	
	\section{$n'$ decay and the $e' - X'$ fluid}
	\label{sec:decay}
	
In this section, we argue that: (i) the assumption of high degree of degeneracy between the $n$ and $n'$,  supplemented  by the assumption that $n'$, like the neutron, is made of three mirror quarks $u'd'd'$;  (ii) mirror quarks have strong, weak and electromagnetic gauge interactions, which are similar to those in  the ordinary  sector. This will imply that
\begin{itemize}
  \item The mirror gauge couplings as well as the mirror quark masses are almost equal to the corresponding SM values, with a high degree of precision.
  \item
  The existence of $(\nu',e')$, with the mirror electron mass $m_{e'}$ also nearly equal to the electron mass $m_e$.
  \item The $p'$ mass must be below the $n'$ mass.
  \item Since $m_{e'} \simeq m_e$ with very high precision, if the  $\nu'_e$ mass is small, 
		there will be  $n'$ beta decay like the familiar neutron with the same rate.
	\end{itemize}
	
	To elaborate a bit on this  argument: clearly if the  quark and mirror quark  masses were different, we would not expect the $n$ and $n'$ masses to be so nearly equal. Similarly, if the gauge couplings of the various ordinary and mirror quarks were not nearly equal, the $n$ and $n'$ masses should receive contributions from color and  mirror color  couplings $\alpha_c$ and $ \alpha'_{c}$ (or the corresponding QCD scales $\Lambda_{\rm QCD}$ and $\Lambda'_{\rm QCD}$), which would upset the required high degree of degeneracy. This then suggests that the proton and its mirror partner $p'$ will also be highly degenerate.  Similarly, the matching of electromagnetic self energy, i.e.
	\begin{eqnarray}
		\Delta m_{\rm EM} \sim \alpha \Lambda_{\rm QCD} \sim {\rm  MeV} \,,
	\end{eqnarray}
	forces a highly precise equality of charge and mirror charge, where $\Delta m_{\rm EM}$ stands for the electromagnetic energy of the quarks, {and $\alpha$ is he fine-structure constant}.
	
	To avoid a mirror photon mass $m_{\gamma'}$ from affecting the mirror electromagnetic self energy above, we need a strict upper bound on the mass of the mirror photon: 
	\begin{eqnarray}
		m_{\gamma'} \leq \sqrt{4\pi \alpha^{-1} \delta_{nn'} \Lambda_{\rm QCD}} \sim 10^{-3} \, {\rm eV} \,.
	\end{eqnarray}
	For simplicity we adopt a massless mirror photon, leading  to the milli-charged scenario~\cite{Holdom:1985ag,Glashow:1985ud}. In this scenario the massless photons organize so as to have one coupled exclusively to the ordinary leptons and quarks. The mirror charged fermions couple mainly to the mirror photon, but carry an ordinary  ``millicharge'' $\epsilon e$, with  $e$ denoting  the electron's charge. The weak radiative corrections imply equality of masses and couplings of ordinary and mirror gauge bosons.
	
	We note parenthetically that putting $m_n \simeq m_{n'}$ with all these assumptions together essentially implies an almost exact mirror  model. There is, however, one more caveat that we have to address. In exact mirror models, there are three extra light neutrinos and the mirror photon. To bring about consistency between three extra neutrinos and an extra photon contributing to the energy density in the Big Bang Nucleosynthesis (BBN) epoch  of the universe  with the Planck data~\cite{Planck:2018nkj}, we require
	that there be asymmetric inflation implemented~\cite{Berezhiani:1995am}. This will remove the above BBN problem by lowering the reheating temperature in the mirror sector by a factor of three. This will then dilute the impact of the extra mirror neutrinos and the mirror photon on BBN, thus restoring consistency. 
	Since asymmetric inflation changes the dynamics of the mirror sector compared to the visible sector, this has the impact on particle physics
	parameters of the model such as $\delta_{nn'}$. It has been shown that this mirror symmetry breaking  effect  can be made very mild and compatible with $m_n \simeq m_{n'}$ in a suitably chosen inflation  model~\cite{Mohapatra:2017lqw, Babu:2021mjg}. 

\begin{figure*}[!t]
  \includegraphics[width=0.75\textwidth]{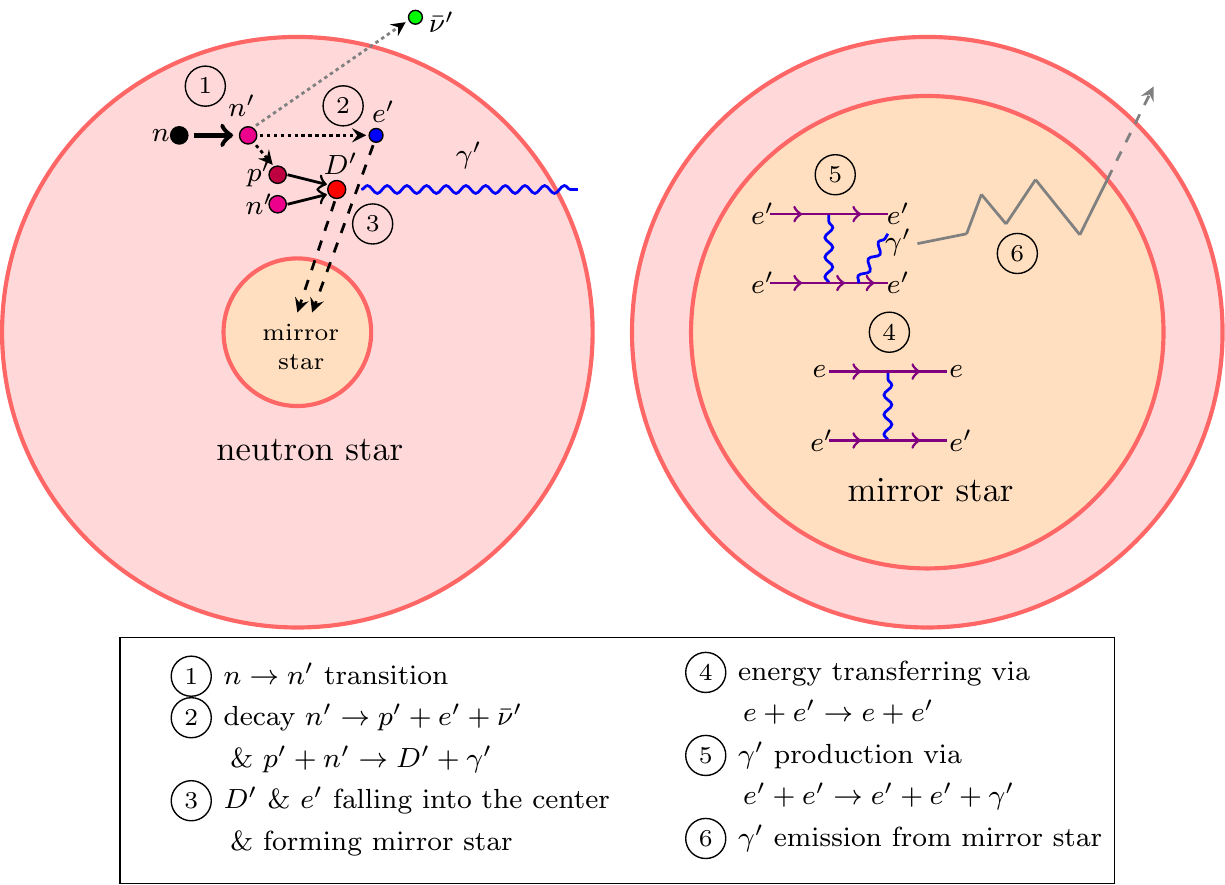}
  \caption{A schematic depiction of what happens after the $n-n'$ transition takes place in a NS. The formation of heavier nuclei such as $_2^4$He$'$ is not shown in this figure, which will not affect the main results in this paper. In the right panel we zoom in the ``mirror star'' region in the left panel. }
  \label{fig:1}
\end{figure*}

	The $\beta$ decay of $n'$ proceeds in the same manner as just stated, and will have the same rate of $\sim (880 \; \rm sec)^{-1}$ as ordinary neutron decay in vacuum, so long as the Fermi energy of the electron is much smaller than the $Q$ value of 0.7 MeV of the $\beta$ decay. If there is a mirror weak interaction similar to the familiar weak interaction, the $\beta'$ decay is~\cite{Berezhiani:2020zck, Berezhiani:2021src}
	\begin{eqnarray}
		n'\to p' +e' +\bar{\nu}'_{e} \,.
	\end{eqnarray}
	The only caveat could be that the $\bar{\nu}'_e$ could be heavier without affecting the $n-n'$ mass degeneracy.
	The new process is depicted schematically in Fig.~\ref{fig:1}. The $p'$s, like the $n'$s, are gravitationally bound to the NS, and local mirror charge neutrality forces $n_{e'}(r) = n_{p'}(r)$ at all $r < R$.
	The mirror neutrons and mirror protons slow down and form mirror deuterons $D'$, since the process
	\begin{eqnarray}
		p'+n'\to D'+\gamma' \,,
	\end{eqnarray}
is faster than the inverse beta decay $ e'+p' \to n' +\nu'$. As in the Sun and other lighter ordinary stars, the following nuclear fusion will be dominated by the mirror proton-proton chain~\cite{Adelberger:2010qa}. For instance, the mirror deuteron $D'$ will undergo further fusion with $p'$ to form $_2^3$He$'$, and two $_2^3$He$'$ will generate $_2^4$He$'$ while releasing two $p'$ and some energy. In such processes, a small fraction of $_3^7$Li$'$ and $_4^7$Be$'$ may also be generated. However, no matter what nuclear processes happen in the mirror sector, charge neutrality holds locally inside the mirror star, i.e. $n_{e'}(r) = \sum_{X'} Z' n'_{}(_{Z'}^{A'} X',\, r)$, with $n'_{}(_{Z'}^{A'} X',\, r)$ the number density of the mirror nucleus $X'$ with $Z'$ $p'$ and $(A'-Z')$ $n'$ as function of radius $r$.

	\section{The profile of  $n_{e'}$}
	\label{sec:profile}
	
	In this section we write down the equations governing the hydrostatic structure of the $e'-X'$ sphere, and then solve them. The mirror electrons $e'$ and mirror nuclei $X'$ constitute a fluid that is supported against the gravity of the ordinary NS by degenerate pressure, which is dominated by that of the $e'$. The mass density of the fluid is dominated by the mirror nuclei $X'$. For simplicity we assume in this section the nuclei in the mirror star are mostly $_2^4$He$'$. Then charge neutrality requires that $n_{e'} (r) \simeq 2n_{_2^4 \text{He}'} (r)$. Since the number density of $_2^4 {\rm He}'$ is half of that of $D'$, the number of $e'$ will not change and the mass density will be the same as the case of $e'-D'$ fluid.


	The $e'$ pressure for a given Fermi momentum $p_F$ is
	\begin{eqnarray}
		P_{e'} = \frac{8\pi}{3 m_{e'} \hbar^3}\int_0^{p_F} {\rm d}p \frac{p^4}{\sqrt{1 +({p/m_{e'} c)^2} }} \,,
	\end{eqnarray}
	and the  hydrostatic equation is
	\begin{eqnarray}
		\label{eqn:hydrostatic}
		\frac{\partial}{\partial r}P_{e'}(r) = - \rho(r) g (r) \,,
	\end{eqnarray}
where the mass density $\rho(r)\simeq n_{e'} (r)m_{_2^4 \text{He}'}/2$.
For the small radii considered, the gravitational acceleration can be approximated by
	\begin{eqnarray}
		g(r) = \frac{G_N M(r)}{r^2} = \frac{4\pi}{3}G_N \rho_0 r \,,
	\end{eqnarray}
	where $G_N$ is the Newtonian constant of gravitation,  $\rho_0 = 10^{15} \, {\rm gr}\, {\rm cm}^{-3}$ is the center mass density of the NS. For $r< 2$ km the density is almost a constant. Moreover, for such small radii,   the general relativistic modifications of the hydrostatic equation are  very small  (a relative correction of $\sim 10^{-3}$). As we shall see, this enables an analytic solution of the hydrostatic equation.
	
	We thus obtain
	\begin{eqnarray}
		&& \frac{8\pi}{3 m_{e'} \hbar^3}\frac{p_F^4}{\sqrt{1 +({{p_F}/{m_{e'} c})^2}}}  \frac{\partial}{\partial r}p_F(r) \nonumber \\
=&&		-   \frac{2\pi}{3}G_N \rho_0  m_{_2^4 \text{He}'}  n_{e'} (r)  r \,.
	\end{eqnarray}
	Substituting in 
	\begin{eqnarray}
		\label{eqn:ne_r}
		n_{e'}(r)= \frac{8\pi}{ 3}\left(\frac{p_F(r)}{\hbar}\right)^3 \,,
	\end{eqnarray}
	and introducing the  dimensionless  parameter $X_F \equiv {p_F}/{m_{e'} c}$,
	we can obtain a very simple equation for $X_F$: 
	\begin{eqnarray}
		\frac{X_F}{ \sqrt{1+ X_F^2} } \frac{{\rm d}}{{\rm d}r}X_F(r)= -    \frac{2\pi  G_N \rho_0 m_{_2^4 \text{He}'}}{3 m_{e'} c^2} r
		= - \frac{r}{r_0^2} \,,
	\end{eqnarray}
	with
	\begin{eqnarray}
		r_0 = \sqrt{ \frac{3m_{e'} c^2}{2\pi  G_N \rho_0 m_{_2^4 \text{He}'}} } \simeq 0.296 \ {\rm km} \,.
	\end{eqnarray}
	The solution is
	\begin{eqnarray}
		\sqrt{  1+X^2_F(0) } - \sqrt{ 1+X^2_F(r) } = \frac{r^2}{2 r_0^2} \,,
	\end{eqnarray}
	leading to 
	\begin{eqnarray}
		\label{eqn:XFr}
		X_F(r)= \left[ \left( \sqrt{X_F^2(0)+1} -  \frac{r^2}{2r_0^2} \right)^2 - 1 \right]^{1/2} \,.
	\end{eqnarray}
	
\begin{figure}[!t]
		\centerline{\includegraphics[scale=0.65]{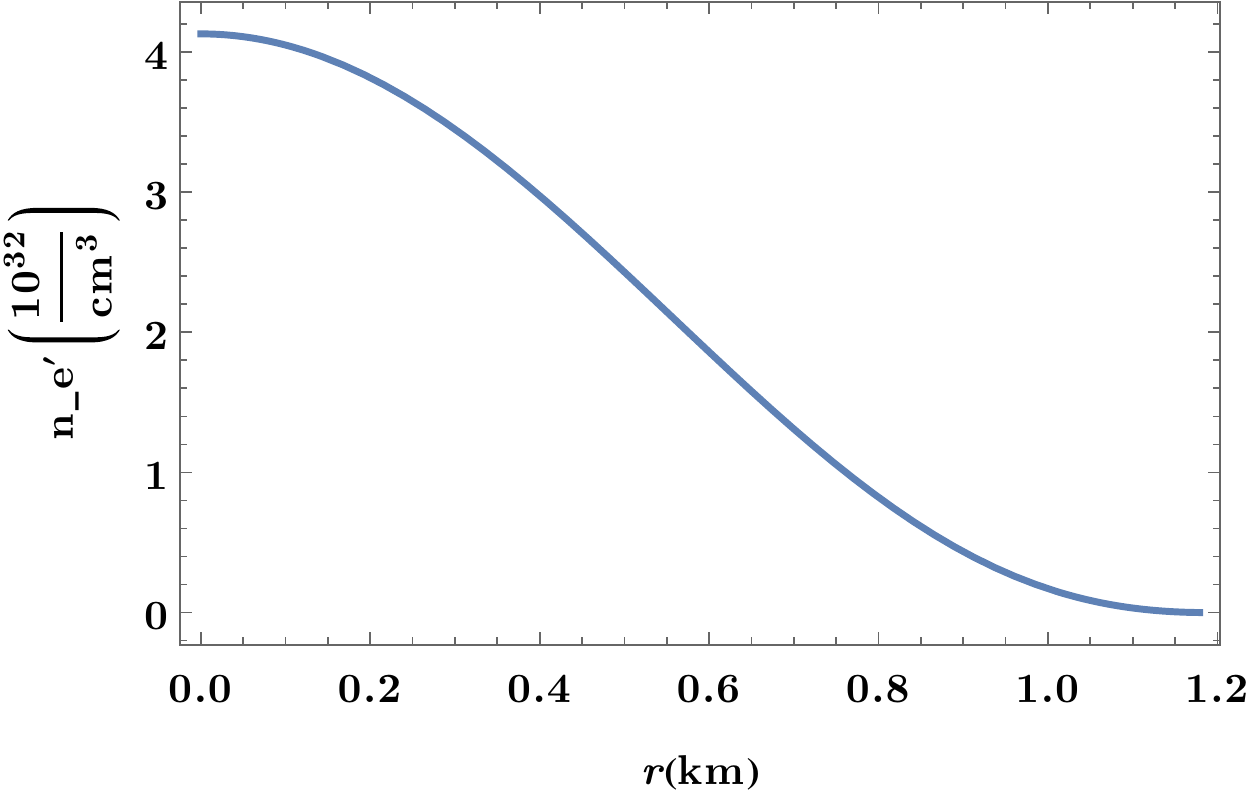}}
		\caption{$n_{e'} $ as function of $r$.}
		\label{fig:2}
	\end{figure}

	Then from Eq.~(\ref{eqn:ne_r}), we can obtain the solution for $n_{e'}(r)$:
	\begin{eqnarray}
		n_{e'}(r) &=& \frac{8\pi}{3m_{e'}^3 c^3 \hbar^3} \left[ \left( \sqrt{X_F^2(0)+1} -  \frac{r^2}{2r_0^2} \right)^2 - 1 \right]^{3/2} \,. \nonumber \\ &&
	\end{eqnarray}
	The  number of the $e'$ up to the radius $r$ is
	\begin{eqnarray}
		{\cal N}_{e'}(r)= \int_0^{r}4\pi n_{e'}(x )x^2 {\rm d}x \,.
	\end{eqnarray}
	The fluid is confined inside a sphere with radius $R_c$  so that  $n_{e'}( R_c) =0$. Once $X_F(0)$ is given, $R_c$, $n_{e'}(r)$ and the total number ${\cal N}_{e'} \simeq 2{\cal N}_{_2^4 \text{He}'}$ are determined by pure numbers and fundamental constants. This  resembles the case  of the Chandrasekhar mass. In order to find the value of $X_F(0)$, we solve the structure equation  iterating the value of $X_F(0)$ until the constraint ${\cal N}_{e'}(R_c)= 5\times 10^{47}$ is satisfied, so that the total number ${\cal N}_{e'} \simeq 2{\cal N}_{_2^4 \text{He}'}$
	is half of the total $n'$ generated. We obtain $X_F(0) = 8.9$, implying  $E_F( 0)= 4$ MeV and $R_c= 1.18$~km.  The mirror electron density $n_{e'}$ as function of $r$ is shown in Fig.~\ref{fig:2}.
	
	\section{Energy drain to the mirror fluid and  neutron star  cooling}
	\label{sec:energy}
	
	Finally we are ready to face the central question of this work: Can the $X'$s, $e'$s and $\gamma'$ provide a fast cooling route for the oldest and coldest NSs, quenching the photonic signals and evading the resulting strict bounds on $\epsilon_{nn'}$? Having even a tiny fraction of weakly interacting particles inside a star can dramatically modify its thermal properties. Recall, for example, that the hypothetical $\sim 5$ GeV DM particles collecting  inside the sun over its age to  $10^{-12}$ level can cool the solar core and was considered as a resolution of the early solar neutrino anomaly~\cite{Press:1985ug,Gilliland:1986}.  The value of the solar luminosity has been recently used to bound a possible interaction between mirror DM and the ordinary solar matter~\cite{Michaely:2019xpz}. In a similar manner, we argue that the additional $10^{-9} $ fraction of mirror
	particles can strongly affect the cooling of old NSs. This is also suggested by having the energy generated matching the production of the $n'$ and ensuing $p'$ and $e'$ mirror particles.
	
	Let us first recap the steady state picture envisioned in deriving the strict bound by using
	the electromagnetic luminosity ${\cal L} = {\rm d}W/{{\rm d}t}$ of the NS. A key point is that the rate of $n\to n'$ transition is constant and independent of any thermal or other variations (except for stopping when the mixed star forms, which happens after many Hubble times for the small values of $\epsilon_{nn'}$ considered). The $\sim 50\%$  of the heat generated which
	resides in the SM component is then radiated via a fixed black body luminosity~\cite{McKeen:2021jbh}.
	Having all the mirror particles segregated in a ``core region'' comprising $\sim 0.1\%$ of the star
	volume would have seemed  to minimize  their ability to intercept and impede ordinary heat emission and photon
	radiation from the mirror free, large outer region. 
	This, in turn, would have suggested only minor luminosity reduction and no relaxing of the bounds on $\epsilon_{nn'}$. However, a more careful scrutiny shows that this simplistic argument is misleading.
	
	The energy emission from the core will be dominated by the radiation of mirror photons, while the heat is continuously transferred from the normal sector to the mirror sector by scatterings of the normal and mirror electrons in the core region. For sufficiently large millicharge $\epsilon$, the heat emission rate from the mirror particles
	may overtake the normal emission rate from the external surface by an appreciable factor. The ordinary photonic energy may then account only for a small part of the energy generated inside the star. Furthermore, the cumulative effect of this over most of the stellar
	history will reduce its heat content and push the internal and external surface temperatures to zero, quenching the photonic emission and destroying the steady state model envisioned.
	
	Thanks to the mutual mirror electromagnetic scattering of the mirror particles inside the core region and attendant emission of the fast escaping mirror photons, the time required for their cooling off
	and equilibrating at a temperature $T'$ is very short on typical thermal timescale of  $t_{\rm thermal} = W^*/({\rm d}W/{{\rm d}t})$, where $W^*=Q^*$ is the total heat content of the star.
	Using Eqs.~(\ref{eqn:LNS}) and (\ref{eqn:Qstar}) we find $t_{\rm thermal} \sim 3\times10^{15}$ sec,  which happens to be close to the age of the star.
	
	Since the emission of heat from the mirror sector is much faster than heat transfer between the sectors, any amount of heat in the mirror sector will be emitted rather than go back to the normal sector, which also implies that
	\begin{eqnarray}
		T' \leq T \,.
	\end{eqnarray}
	To avoid detailed discussion at the particle scattering level, we first view the core region as a black body for the {\it mirror} photons with temperature $T'$,  as indeed it absorbs any such photon falling on it .  The surface of area $4\pi R_c^2$ of the inner "core region" serves effectively as an additional boundary, through which the heat in the normal component of the surrounding star can be emitted.
	The mirror electrons in the core will then radiate their heat content to the outside with the
	rate of black body luminosity:
	\begin{eqnarray}
		{\cal L}' = 4\pi \sigma_{\rm SB} R_c^2 T'^4\ .
	\end{eqnarray}
	Relative to the internal core region surface $4\pi R_c^2$, the stellar surface is larger --  by roughly a factor of  100. However the thermal blanket makes the internal temperature about hundred-fold bigger than the surface temperature.
	Thanks to the possibility that  $T^{'4} \geq 10^8 T_s^4$, even if we keep $T' < T$ to make $e\to e'$ energy transfers more than the reverse transfer, we can still, in principle, have the rate of mirror photon emission almost six orders of magnitude bigger than that of the
	ordinary photons, so long as  $R_c\geq 1$ km.
	
	However, to verify that this indeed happens, we need to check how many $e-e'$ collisions occur per second (which we denote by $\dot{\cal N}_{\rm col}$)
	between the ${\cal N}_{e}( r<R_c)\sim  10^{38}R^3_c \ {\rm cm}^{-3}$ electrons in the core region and the  ${\cal N}_{e'} \sim 5 \times 10^{47}$ mirror electrons. If the total energy transferred per second via these collisions from the ordinary to mirror electrons much exceeds the stellar luminosity, namely the inequality
	\begin{eqnarray}
		\label{eqn:inequality}
		\dot {\cal   N}_{\rm col} \Delta E \sim \dot{\cal N}_{\rm col} \Delta T \gg {\cal L}_{\rm NS} \sim 2\times 10^{36} ~ {\rm keV}~{\rm sec}^{-1}
	\end{eqnarray}
	holds, then the mirror luminosity dominates and the scenario envisioned in deriving the strict upper bounds on $\epsilon_{nn'}$ becomes inoperative. On the other hand, if
	the inequality in Eq~(\ref{eqn:inequality})  is (strongly) reversed, then the above scenario involving the $\beta$ decay of the mirror neutron will be irrelevant.
	
For the average energy transfer of $\Delta T \sim 0.35$ keV, Eq.~(\ref{eqn:inequality}) becomes $\dot {\cal N}_{\rm col} \geq10^{37}$ sec$^{-1}$. Each electron and also each mirror electron move with the speed of light $c$. Then we can express $\dot{\cal N}_{\rm cal}$ with energy transfer of $\sim$0.35 keV in a manner, which is symmetric between the ordinary and mirror sectors:
	\begin{eqnarray}
		\dot{\cal N}_{\rm cal} =  \frac{c f f'{\cal N}_e (r< R_c) {\cal N}_{e'} \sigma_{ee'} } {(4\pi/3) R_c^3}  \,,
	\end{eqnarray}
	where $ f=kT/E_F\sim 10^{-5}$ and $f' = kT'/{E'_F} \sim 10^{-4}$ are the fractions of the ``active'' electrons and mirror electrons, respectively. For $R_c=1.2$ km,
	collecting all these factors, the condition $\dot{\cal N}_c \gg 10^{37}$ sec$^{-1}$ translates into the following requirement on the $e-e'$ scattering cross section:
	\begin{eqnarray}
		\sigma _{ee'} \simeq \epsilon^2 \sigma_{ee} \gg 10^{-50} ~{\rm cm}^{2}  \,,
	\end{eqnarray}
where $\sigma _{ee}$ is the standard Rutherford scattering cross section of electrons in the same kinematic configuration. For the formula for $\sigma_{ee'}$, see Appendix~\ref{sec:xs}. The important absence of the $u$-channel diagram in the calculation of $e-e'$ scattering and some analogue processes have been emphasized by Ref.~\cite{Girmohanta:2022dog}.
Including only the Feynman diagram for the $t$-channel photon exchange, the cross section $\sigma_{ee'}$ is calculated by having the relativistic $e$ and $e'$ with energies $E_F \simeq 10 E'_F\simeq 35$  MeV  collide at random relative direction in the laboratory frame and transferring an energy of $T \sim  0.35$ keV between them. Using a plasmon mass as the cutoff, we estimate this cross section to be $\sigma_{ee'} \simeq {4\pi \alpha^2\epsilon^2}/{E_FT} \simeq 10^{-23} \epsilon^2$ cm$^2$ (see Appendix~\ref{sec:xs} for details), which leaves us with the rather weak, easy to satisfy requirement
	\begin{eqnarray}
		\label{eqn:epsilon}
		\epsilon^2 \gg 10^{-27}  \,.
	\end{eqnarray}

There are laboratory limits on the mixing parameter from positronium decays, which result in $\epsilon < 5\times10^{-8}$~\cite{Vigo:2019bou}. The cosmological bounds from the BBN are yet stronger: for the case of exact mirror parity, the limit is $\epsilon < 3\times 10^{-10}$, for the case of asymmetric mirror sector, the limit can be much weaker~\cite{Berezhiani:2008gi}. This still leaves enough margin for satisfying  Eq.~(\ref{eqn:epsilon}). The strongest upper bound $\epsilon\leq 10^{-12}$~\cite{Vogel:2013raa} do not apply here, as in mirror models the DM is made of neutral objects such as the composite mirror Hydrogen, deuteron or Helium.
Interestingly, these limits remain in the range of interest for a direct detection of mirror nuclei as DM via Rutherford scattering mediated by this kinetic mixing~\cite{Cerulli:2017jzz}.
A full list of the laboratory, astrophysical and cosmological limits and future prospects on mirror photon mass and its kinetic mixing with the SM photon can be found e.g. in Ref.~\cite{Caputo:2021eaa}. The parameter space of $10^{-27} \lesssim \epsilon^2 \lesssim 10^{-19}$ of interest in this paper can be probed by ALPHA~\cite{Gelmini:2020kcu}, BREAD~\cite{BREAD:2021tpx}, DARK E-field~\cite{Godfrey:2021tvs}, LAMPOST~\cite{Baryakhtar:2018doz}, MADMAX~\cite{Gelmini:2020kcu} and SuperCDMS~\cite{Bloch:2016sjj}.
	

\section{Comments}
\label{sec:comments}
	
Here are some comments:
\begin{enumerate}
  \item  This alternative dominant channel of energy emission via an unobserved hidden channel undermines the ability to put upper bounds on the $n \to n'$ transition rate. Even assuming that this is the only source of internal energy, it evades the strict upper bounds on $\epsilon_{nn'}$ claimed. The energy transfer happens for a wide range of $\epsilon$, which are definitely allowed and possibly even favored within mirror models. Furthermore, this motivates the fascinating transition of NS to a mixed normal-mirror NS in the first place.
		
      Even though the photonic cooling of UCNS  is not a reliable way to set bound on the $n \to n'$ transition rate for the case of near exact mirror symmetry and slow $n \to n'$ transition, there are situations when it works: e.g. (i) we could have a near exact mirror symmetry but the millicharge of the mirror fermions $\epsilon \lesssim 10^{-13}$ or, (ii) an asymmetric mirror model with $m_{p'} \geq m_{n'}$ where $n'$ is the DM of the universe, so that $\beta$ decay of mirror neutron is forbidden. It can also work in 		other dark baryon contexts, such as those suggested in connection with the neutron lifetime anomaly~\cite{McKeen:2015cuz}.
	
  \item Actually one can compare the energy loss of NSs due to the energy transition mechanism in this paper with that due to radiation of photons. If the former is comparable to the latter, or a few time larger, the NS limit on the $n-n'$ mixing parameter $\epsilon_{nn'}$ will also be weakened, at least to some extent. However, the NS limits in such scenarios will depends on how much energy is transferred to the mirror sector, which goes beyond the main scope of this paper.

		
		\item An advantage of the heating up argument as compared with the orbital period stability method~\cite{Goldman:2019dbq} is that: in principle in the former one can use all pulsars, whereas for the other one using orbital period data requires binary pulsars~\cite{Goldman:2009th, Goldman:2019dbq}.
		
		\item Unfortunately, unlike the misquote in Ref.~\cite{McKeen:2021jbh}, the spinning period changes of single pulsars -- which, as part of the ambitious nano-gravity project, have been determined in many cases with stunning accuracy -- {\it cannot} be used, as it is affected by relatively large and incalculable changes due to magnetic braking etc. This is the reason why  binary pulsars were used in Refs.~\cite{Goldman:2009th, Goldman:2019dbq}.
		
		\item As  PSR J2144$-$3933 used in Ref.~\cite{McKeen:2021jbh} seems to be a unique outlier in many respects, one should reserve judgement until more UCNSs of the same or lower temperature/luminosity are discovered. The feature that most clearly distinguishes it from the other 55 pulsars included in Ref.~\cite{Yakovlev:2004iq} is the luminosity, which is a factor of 30 lower than that of the next low luminosity quartet of pulsars shown in  Fig~1.  of Ref.~\cite{McKeen:2021jbh}. Also its spin period of  $8.5$ seconds, the longest among observed radio pulsars, combined with a small magnetic field, make it a very weak pulsar. Indeed we happen to see it just because it is relatively close. It is also worth mentioning that, with three other pulsars, it was not included in the tables and figures of the exhaustive reviews of Refs.~\cite{Yakovlev:2004iq, Potekhin:2020ttj}, as these pulsars do not contribute to the understanding of the heat conduction and cooling mechanism of old cool pulsars. Finally, the pulsar PSR J2144$-$3933 is also rather unique in terms of its orbital parameters, which were studied using very high precision  observations~\cite{Deller:2009tc,Guillot:2019ugf}. The tangential velocity components  in and out of the plane of the galaxy are large, suggesting that along almost all its trajectory it is at much larger distances, where finding its temperature and luminosity would be impractical. Since we do not expect any correlations between the orbital and thermal parameters, it would then suggest that many more nearby pulsars are just as cold or colder.

\item The energy transfer mechanism in this paper can be applied equally to other NSs, e.g. those in Ref.~\cite{Yakovlev:2004iq} although the surface temperatures of these stars are higher and the luminosities are larger than PSR J2144$-$3933. However, the limits of these stars on $\epsilon_{nn'}$ is to some extent weaker than that from PSR J2144$-$3933, the relaxation of these limits due to the emission of $\gamma'$ will depend on the stellar sizes, temperatures etc. More generally, the transfer mechanism may also cause faster cooling of younger neutron stars, thus affecting significantly the emission of neutrino pairs or gravitational waves. Then this mechanism can be constrained by the corresponding astrophysical observations. However, this is beyond the main scope of this paper.
\end{enumerate}
	
	We close this section with the following comments:  it is clear that a milli-charge portal connection between the ordinary and mirror sectors is important for our conclusion, and if there is no such connection, the strict bounds on $\epsilon_{nn'}$ derived in Ref.~\cite{McKeen:2021jbh} will be valid. Even so, there is the question as to whether on the basis of a single UCNS the PSR J2144$-$3933 one should give up on the efforts to measure $\epsilon_{nn'}$ as small as $10^{-17}$ eV. We believe not, until more similar NSs are discovered.
	
	\section{Role of accretion}
	\label{sec:accretion}
	
	We now consider the effect of accretion on the NS luminosity. Using the heating of UCNS  to detect dissipative DM has been suggested by many authors. A key point is the accretion of mutually non-interacting DM particles, where the gravitational focusing is controlled by angular momentum conservation along the trajectory of each particle, and the accretion radius is:
	\begin{eqnarray}
		R_a =\frac { v_{\rm esc} }{{v_{\infty}}} R= \kappa_v R \,,
	\end{eqnarray}
	where $v_{\rm esc} \sim 0.4 c$ is the escape velocity from the NS, $v_{\infty} \sim v_{\rm virial} \sim 10^{-3} c$ is the much smaller velocity at infinity of the particles, and $\kappa_v = v_{\rm esc}/v_\infty$. On the other hand, when the particles accreted are strongly mutually interacting and in particular also dissipative, then angular momentum is not conserved and only energetics can be used. The star then accretes from distances up to the   Bondi-Hoyle-Lyttleton (BHL)  accretion radius
	\begin{eqnarray}
		R_{\rm BHL}\sim \frac{v^2_{\rm esc}}{v^2_{\infty}} R  = \kappa_v^2 R \,,
	\end{eqnarray}
	and the corresponding effective areas are $\kappa_v^2\sim 10^5$ times larger than in the non-interacting case. Every accreted particle which does stay in the star contributes $\sim 20\%$  of its rest mass to the energy, which is na\"{i}vely expected to be radiated via photons. For the particle to stay in the NS, it should lose a fraction $\kappa_v^{-2} \sim 10^{-5}$ of its energy in the first collision inside the star. For having such a collision, the optical depth is required to be\footnote{Having the interactions between the ordinary and dark sectors mediated via dark photon may be problematic,
as most of the energy could be emitted via dark photons.}
	\begin{eqnarray}
		\label{eqn:opticaldepth}
		n_n R\sigma \geq 1 \,,
	\end{eqnarray}
which is easy to obtain, where $n_n \sim 10^{39}$ cm$^{-3}$ is the number density of neutron in the NS, and $\sigma\geq 10^{-45} \; {\rm cm}^2$ is the cross section between the infalling particle and the neutron in the NS.

Accretion of interstellar gas onto the NS surface    will liberate energy, which is a fraction  $1 - \sqrt{ 1- 2 G_N M/c^2 R}$ of the mass accreted. For $M= 1.5 M_{\odot}$ (with $M_\odot$ the solar mass) and $R= 12$ km it is $\sim$0.2. The latter thermalizes and is radiated from the surface. Let us consider accretion from the ISM onto a NS with mass $M_{\rm NS}$, radius $R$ and velocity $v_{\infty}$ relative to the ISM. The mass accretion rate is $ \dot{M} = \pi v_{\infty} R_{\rm BHL}^2 \rho$, where
	$\rho$ is the mass density of the accreted ISM material.
	The pulsar {PSR J2144$-$3933} under consideration is  above the mid plane of galactic disk, so we adopt a Hydrogen number density of $ 0.1 \ {\rm cm}^{-3}$. With  $M_{\rm NS} = 1.5 M_{\odot}$ and $R= 12$ km, the transverse velocity of the pulsar turns out to be $132$ km/s~\cite{Deller:2009tc}, implying that $v_{\infty}>132$ km/s. Considering the low ISM temperature, this is indeed supersonic.
	The corresponding accretion radius is  $R_a< 1.82 \times 10^{12}$ cm, and the mass accretion rate is $\dot{M}< 9 .5\times 10^6$ gr/sec. If it would have been accreted onto the NS, the accretion luminosity would have been $1.6 \times 10^{27}$ erg/sec, implying a surface temperature of $ 3.6 \times 10^4$ K,  which is about the observational  upper limit~\cite{Guillot:2019ugf}.
	
	However, {\it accretion would not occur}, since  the pulsar wind energy density at the accretion radius exceeds by three orders of magnitude the gravitational energy density of the inflowing gas at the accretion radius~\cite{Deller:2009tc,Guillot:2019ugf}. Moreover, even for a weaker pulsar wind, accretion would be prevented due to the strong rotating magnetic field, which will disperse the accreted gas via the propeller mechanism~\cite{Treves:1999ne}. The absence of accretion of ISM gas onto NS  is important for motivating a
	campaign searching for UCNS, as otherwise the above estimate of the BHL accretion may suggest that we will not be able to find many pulsars colder than the UCNS PSR J2144$-$3933.
	

	
	\section{Summary}
	\label{sec:summary}
	
	To summarize  the main results of this paper:  the photonic luminosities of UCNSs do not  necessarily imply robust bounds on $\epsilon_{nn'}$. In particular, they do {\it not} exclude discovery via terrestrial measurements of the tiny $ \epsilon_{nn'}\sim {\cal O}(10^{-17}$ eV). This is achieved via a small extra follow up (by $\sim 10^3$ seconds) of the $n\to n'$ saga in NSs, which is a novel, new scenario to our mind. In this scenario, under the joint effect of the weight of the mirror nuclei and the Fermi energy of the mirror electrons, the mirror nuclei and electrons form a configuration resembling that of a ``mini white dwarf'' inside the NS. A remarkable feature of this configuration is its universality stemming from, and in analogy with, the features of NSs and actual white dwarfs. Within this structure, heat is transferred relatively fast (on characteristic thermal time scales of the NS) from the heat
	reservoir in the normal matter of the NS to the mirror sector, and is radiated via mirror photons.

	\section*{Acknowledgement}
	The work of R.N.M. is supported by the US National Science Foundation grant no. PHY- 1914631. Y.Z. is supported by the National Natural Science Foundation of China under grant No.\  12175039, the 2021 Jiangsu Shuangchuang (Mass Innovation and Entrepreneurship) Talent Program No.\ JSSCBS20210144, and the ``Fundamental Research Funds for the Central Universities''.

	
	\appendix
	\section{$e - e'$ scattering cross section}
	\label{sec:xs}
	
	
	In this appendix, we give the formulae for $e-e'$ scattering cross section. The scattering of $e$ and $e'$ is very similar to the $e^- - e^-$ M\o ller scattering, with the $e-e'$ scattering having only the $t$ channel diagram, since $e$ and $e'$ are not identical particles. The amplitude square for $e-e'$ scattering is given by
	\begin{eqnarray}
		\frac14 |{\cal M}|^2 &=& \frac{2e^2 e^{\prime2} \epsilon^2}{t^2}
		\Big[ s^2 + u^2 -8m_e^2(s+u) + 24m_e^4 \Big] \,, \nonumber \\ &&
	\end{eqnarray}
	with $m_e$ the mass of $e$ and $e'$ (assuming for simplicity the masses of $e$ and $e'$ are the same). In the relativistic limit,
	\begin{eqnarray}
		\frac14 |{\cal M}|^2 &=& 32 \pi^2 \alpha^2 \epsilon^2
		\left[ 1 - \frac{2}{\sin^2\left(\theta/2\right)} + \frac{2}{\sin^4\left(\theta/2\right)} \right] \,, \nonumber \\ &&
	\end{eqnarray}
	with $\theta$ the scattering angle in the center-of-mass frame. Then the differential cross section reads
	\begin{eqnarray}
		\frac{d\sigma}{d\Omega} &=& \frac{1}{4E_1 E_2} \frac{1}{32\pi^2} \frac14 |{\cal M}|^2 \nonumber \\
		&=& \frac{\alpha^2 \epsilon^2}{4E_1 E_2} \left[ 1 - \frac{2}{\sin^2\left(\theta/2\right)} + \frac{2}{\sin^4\left(\theta/2\right)} \right] \,,
	\end{eqnarray}
	where $E_{1,\,2}$ are the energies of $e$ and $e'$ in the initial state in the star frame.
	
	The presence of $\sin(\theta/2)$ in the denominator implies a mostly forward scattering. The expression is divergent for $\theta=0$ and we put the cutoff at the plasmon mass
	in the fluid in our estimate.
	In the dense $e-p$ fluid the (ordinary) photon behave as a plasmon with a mass $m_\gamma$ equal to the plasma frequency $\omega$, i.e. $m_\gamma=\omega$. The $e-e'$ scattering cross section is therefore  proportional to $m_\gamma^{-2} = \omega^{-2}$ instead of $(TT')^{-1} \sim T^{-2}$. In normal metals  with $n\sim10^{24} \, {\rm cm}^{-3}$, the standard expression
	\begin{eqnarray}
		\label{eqn:omega}
		\omega^2 = 4\pi e^2 n_e/m_e
	\end{eqnarray}
	yields a plasma frequency corresponding to an energy of $\sim$15 eV. Having here $n_e \sim 10^{38} \; {\rm cm}^{-3}$, i.e. $10^{13}$ times higher, leads to a modification of $e-e'$ cross section, which is $2\times10^{-11}$ times smaller. This dramatically reduces the range of $\epsilon$, for which the basic constraint of ${\cal N}_{\rm col} > 10^{37}$ is satisfied.
	
	However, the plasma frequency in Eq.~(\ref{eqn:omega}) is invalid here. As mentioned in Section~\ref{sec:conversion}, in the highly degenerate electron fluid only a small fraction $f=T/E_F$ of ``active'' electrons can respond to an external oscillating electric field, much like the fact that only the electrons at the top of the conduction band in metals can freely respond.  In the NS, the electron density $n_e \sim 4k_F^3/{9\pi}$. With the high Fermi energy $E_F \sim 20\ {\rm MeV} \gg m_e$ of the electrons in the NS, the electrons are relativistic, and we can write the density of the relevant ``active'' electrons as:
	\begin{equation}
		n_{e,\,{\rm active}} = f \times \frac{4E_F^3}{9\pi} = \frac{4 E_F^2 T}{9\pi} \,.
	\end{equation}
	Also, the electron mass $m_e$ representing the inertial resistance of the system to oscillating is no longer relevant, and should be replaced by its Fermi energy $E_F$ in Eq.~(\ref{eqn:omega}). Making these two changes in Eq.~(\ref{eqn:omega}), we have
	\begin{equation}
		m_{\gamma}^2 = \omega^2 = \frac{16}{9} E_F T \,.
	\end{equation}
	The $e-e'$ Rutherford cross section will then be reduced by
	\begin{equation}
		\frac{T}{E_F} \sim \frac{0.35 \rm \; keV}{20 \rm \; MeV} \sim 10^{-5} \,,
	\end{equation}
	which still allows ${\cal N}_{\rm col} > 10^{37}$ so long as $\epsilon > 10^{-13}$.

	\bibliography{ref}
	
\end{document}